\documentstyle[makeidx,epsfig,NATO]{crckapb} 
\input epsf.sty
\makeindex 
\begin{opening}
\title{NON-LTE RADIATION PROCESSES: \protect\\
APPLICATION TO THE SOLAR CORONA}

\author{S. COLLIN}
\institute{Observatoire de Paris \\ F--92195 Meudon, France}
\end{opening}

\runningtitle{}
\begin{document}

\begin{abstract}
These lectures are intended to present a simple but relatively complete 
description of the theory needed to understand the formation of lines in  
non-local thermodynamical 
equilibrium (NLTE), without appealing to any previous knowledge except a 
few basics of physics and spectroscopy. After recalling elementary notions 
of 
radiation transfer,
the chapter is focussed on the computation of the level populations, the 
source function, the ionization 
state, and finally the line intensity. 
An application is made to forbidden coronal lines which were observed 
during 
eclipses since decades.
\end{abstract}

\section{Introduction}

Astrophysical media, at least for the fraction we can observe, are 
often 
very dilute compared to those we are used to on Earth.  In the 
interstellar medium for instance, the number density is in average 
one 
atom per cm$^3$, in  HII regions and planetary nebulae it is about 
10$^4$ atoms 
per cm$^3$ and in the solar corona 10$^8$ atoms per cm$^3$, orders of 
magnitudes 
less than those one can get in laboratory experiments. Therefore the 
notion of ``thermodynamical equilibrium" \index{thermodynamical equilibrium} 
does not hold, and quite 
unusual 
phenomena are taking place, such as the emission of intense 
``forbidden 
lines" \index{forbidden line} never observed on Earth. We will in particular show how to compute 
the intensity \index{intensity} of visible forbidden coronal lines,
\index{coronal line} and show how these 
lines, 
and others, can be used to get physical parameters of the solar corona.

\section{Basics of Transfer}

\subsection{PHOTOMETRIC QUANTITIES}
		
We shall first define a few basic photometric quantities. Note that CGS 
units will generally be used.

\begin{figure}
\centerline{\psfig{file=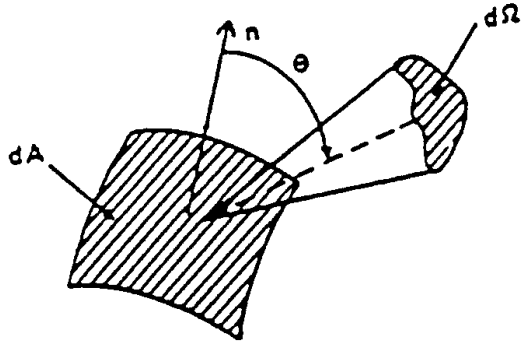,width=0.6\textwidth,angle=0,
clip=}}
\caption{The specific intensity. \index{intensity}}
\label{fig-Inu}
\end{figure}

\noindent$\bullet$\quad  {\bf The specific intensity $I_{\nu}$}:  \index{intensity}		
The energy $dE_{\nu}$ crossing a surface of area $dA$ in the direction of 
the normal, in a solid angle 
$d\Omega$,  during a time 
$dt$, in a frequency interval $d{\nu}$, is (cf. Fig. \ref{fig-Inu}):
\begin{equation}
dE_{\nu}  = I_{\nu}\ d{\nu}\ d\Omega\ dt\ dA.
\label{eq-Inu}
\end{equation}
 It can be expressed in Watt m$^{-2}$ ster$^{-1}$ Hz$^{-1}$, but 
other  units are used in 
the visible or in the X-ray range. It can be also defined per interval 
of 
wavelength, and according to the relation $\lambda\nu=c$ (speed 
of light), one gets 
$I_{\lambda} = I_{\nu} c/\lambda^2$.

	The specific intensity \index{intensity} can be used both for the source
 of radiation or 
for the receptor, and it is constant along a path ray in the vacuum. 

\medskip
\noindent$\bullet$\quad  {\bf The mean intensity  $J_{\nu}$}: 
\index{intensity}
\begin{equation}
J_{\nu}  ={1\over4\pi}\ \int I_{\nu}\  d\Omega 
\label{eq-Jnu}
\end{equation}

\noindent$\bullet$\quad  {\bf The flux $F_{\nu}$}: \index{flux}

\begin{figure}
\centerline{\psfig{file=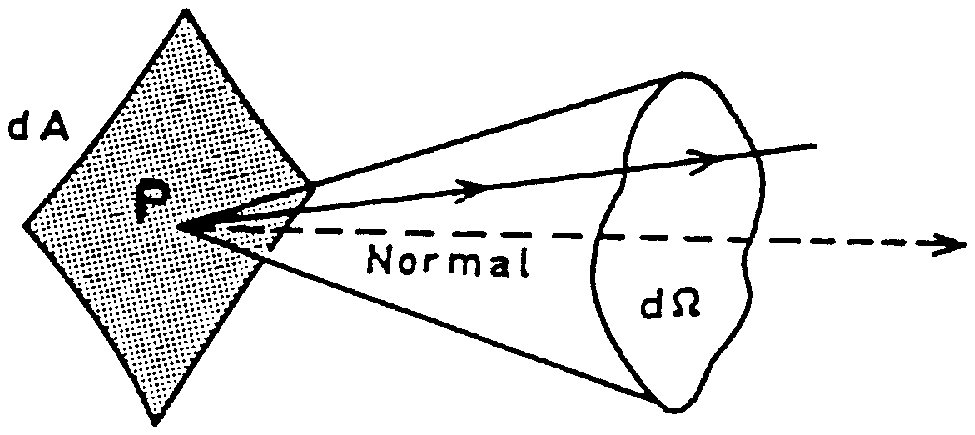,width=0.6\textwidth,angle=0,clip=
}}
\caption{The flux. \index{flux}}
\label{fig-Fnu}
\end{figure}

It is the power crossing 
a unit surface per unit frequency interval, in 
all directions (cf. Fig. \ref{fig-Fnu}):
\begin{equation}
F_{\nu}  = \int I_{\nu} \ \cos\theta\ d\Omega 
\label{eq-Fnu}
\end{equation}
$F_{\nu}$ is generally expressed in Jansky in the radio and far infrared 
range:   1 J = 10$^{-23}$ in CGS, or 
10$^{-26}$   in MKS 
(Watts m$^{-2}$ Hz$^{-1}$).

\begin{figure}
\centerline{\psfig{file=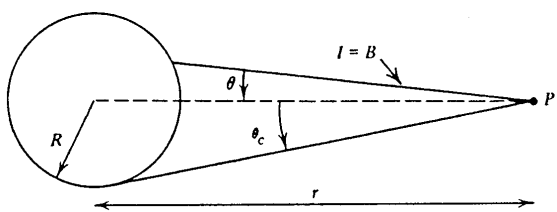,width=0.6\textwidth,angle=0,clip
=}}
\caption{The flux received from a star.\index{flux}}
\label{fig-flux-etoile}
\end{figure}

	As an application, we can compute the flux \index{flux} from a uniform 
sphere 
radiating isotropically (not a very good approximation for a star, 
actually), cf. Fig. \ref{fig-flux-etoile}): 
\begin{equation}
F_{\nu}  = \int I_{\nu}\ \cos\theta\ d\Omega
=  I_{\nu} \int_0^{2\pi}d\phi\int _0^{\theta_c}\sin\theta \ \cos\theta 
\ d\theta,
\label{eq-Fnu*}
\end{equation}
where $\theta_c$ is the angle under which the sphere is seen from 
the observer. If 
$R$ is the radius of the star, and $r$ its distance, one gets $\sin 
\theta_c=R/r$, and 
\begin{equation}
F_{\nu}=\pi I_{\nu} (R/r)^2.
\label{eq-Fnu*bis}
\end{equation}
Note that the flux at the surface of 
the sphere 
is $\pi I_{\nu}$. \index{flux}

\subsection{TRANSFER EQUATION} \index{transfer equation}

When a light ray does not propagate in the vacuum, the specific intensity 
\index{intensity} is 
not 
constant: {\it emission} adds energy, and {\it absorption} removes 
energy. There is also {\it diffusion}, in which the global luminous 
energy  is not changed, 
but  
can be increased in one direction, and decreased  in another. We shall not 
take into 
account the polarization of the radiation, and the possibility of non 
stationary phenomena. 

\subsubsection{Transfer equation in a non diffusive medium}  
\index{transfer equation}

The monochromatic emissivity $\eta_{\nu}$ \index{emissivity} is defined as 
the  power emitted 
per   unit 
solid angle, per unit frequency interval, per unit volume. The 
monochromatic 
absorption coefficient \index{absorption coefficient} $\chi_{\nu}$ 
\index{absorption coefficient} is 
defined 
as follows:  
$\chi_{\nu}I_{\nu}$ is 
the power absorbed per   unit 
solid angle, per unit frequency interval, by 
a slab of unit length normal to the direction of the propagation. Note 
that here  $\chi_{\nu}$ is the inverse of a length. Some people use 
instead the 
``opacity coefficient", \index{opacity} defined per unit mass, and 
the emissivity per 
unit 
mass and not per unit volume. \index{emissivity}
The variation of $I_{\nu}$ on a path length $d\ell$ {\it in the direction 
of 
the light 
ray} is therefore: 
\begin{equation}
dI_{\nu}  =(-\chi_{\nu}I_{\nu}\ +\ \eta_{\nu})d\ell.
\label{eq-dInu}
\end{equation}
Let us define:
\begin{equation}
d\tau_{\nu}  =-\ \chi_{\nu}d\ell \ \ {\rm and} \ \tau_{\nu}  =\int 
-\chi_{\nu}d\ell,
\label{eq-tau}
\end{equation}
$\tau_{\nu}$ is the {\it optical depth}, \index{optical depth} which 
decreases 
towards the observer.

The transfer equation  writes then: 
 \index{transfer equation}
\begin{equation}
{dI_{\nu}\over d\tau_{\nu}}\  =\ I_{\nu}\ -\ S_{\nu},
\label{eq-transfer}
\end{equation}
where $S_{\nu}=\eta_{\nu}/ \chi_{\nu}$ is called the {\it 
source 
function}. \index{source function}

If there are several emission and absorption processes, they have to be 
all taken into account in the source function and in the absorption 
coefficient. For instance, if at a frequency $\nu$ a line is superposed 
onto a continuum, then $S_\nu= S_\nu^{line}+S_\nu^{cont}$, and 
$d\tau_\nu= d\tau_\nu^{line}+d\tau_\nu^{cont}$.

\subsubsection{Transfer equation in a diffusive medium}  \index{transfer equation}

If there is a diffusion process, it must also be
taken into account in the transfer equation.  \index{transfer equation}
 The {\it diffusion 
coefficient} \index{diffusion coefficient} is defined like the absorption 
coefficient: $\sigma_{\nu}I_{\nu}$ is 
the power diffused per   unit 
solid angle, per unit frequency interval, by 
a slab of unit length normal to the direction of the propagation.  Let us 
assume that diffusion is 
coherent (i.e. without any change of frequency), and isotropic. As a 
consequence there is a corresponding emission coefficient, which is equal 
to 
$\sigma_{\nu}J_{\nu}$, and the transfer equation becomes: \index{transfer equation}
\begin{equation}
dI_{\nu}  =[(-\chi_{\nu}-\sigma_{\nu})I_{\nu}\ +\ 
(\sigma_{\nu}J_{\nu}+\eta_{\nu})]d\ell.
\label{eq-dInu-diff}
\end{equation}
 If one defines now an {\it extinction coefficient}, $\tau_\nu^{tot}=
\int d\tau_\nu^{tot}$ with $d\tau_\nu^{tot}=-(\chi_{\nu}+\sigma_{\nu}) 
d\ell$, one gets for the transfer equation in presence of diffusion:
 \index{transfer equation}
\begin{equation}
{dI_{\nu}\over d\tau_{\nu}^{tot}}\  =\ I_{\nu}\ -\ S_{\nu}^{tot},
\label{eq-transfer-diff}
\end{equation}
where $S_{\nu}^{tot}=(\sigma_{\nu}J_{\nu}+\eta_{\nu})/
(\chi_{\nu}+\sigma_{\nu})$. Although it is formally similar to Eq. 
\ref{eq-transfer}, it differs in that the intensity appears directly in 
the 
second term, so it is an integro-differential equation. Actually, the 
diffusion process is a probabilistic one, similar to a random walk, and 
one 
can show that the distance that a photon will travel before being 
absorbed is equal to $[\chi_\nu (\chi_\nu+\sigma_\nu)]^{-1/2}$, while in 
the 
case of pure absorption it is equal to $\chi_\nu^{-1}$. 

In a purely diffusing medium, the transfer equation writes:
 \index{transfer equation}
\begin{equation}
{dI_{\nu}\over d\tau_{\nu}^{dif}}\  =\ I_{\nu}\ -\ J_{\nu},
\label{eq-transfer-diffbis}
\end{equation}

In the corona, the diffusion process is due to Thomson scattering by free 
electrons. As a first approximation it can be considered as a coherent and 
isotropic process. The diffusion coefficient  \index{diffusion coefficient}
$\sigma_\nu$ does not depend on 
frequency,
 and is equal to $\sigma_T N_e$, where $N_e$ is the 
number of electrons per unit volume, and $\sigma_T$ is the Thomson cross 
section, equal to 6.65 $10^{-25}$ cm$^2$. This process is very important 
for the continuum in the visible range, as it is responsible of the 
emission of the K corona. As a consequence one can consider that the corona 
is a purely 
diffusive medium for the continuum in the visible range. On the other hand we will 
show that 
diffusion 
is negligible in the transfer of the coronal lines, \index{coronal line} on which we 
will focus later on, so 
we will not consider it in the following sections.

\subsection{APPLICATION TO A PLANE-PARALLEL MEDIUM}

\begin{figure}
\centerline{\psfig{file=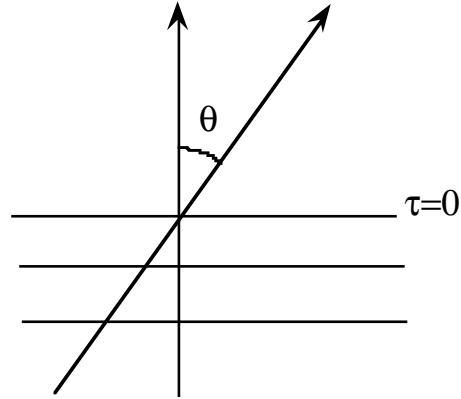,width=0.5\textwidth,angle=0,clip
=}}
\caption{The plane-parallel approximation.}
\label{fig-plan-parallel}
\end{figure}

A common utilisation of the transfer equation  \index{transfer equation}
concerns a stratified 
medium, where all physical quantities are constant on infinite 
parallel planes. 
It is 
the usual approximation made for stellar atmospheres (cf. Fig. 
\ref{fig-plan-parallel}). Often interstellar clouds or 
diffuse nebulae are considered also as plane parallel media. The 
optical 
depth is then defined in the direction of the normal, and the transfer 
equation becomes,   for a light ray which makes an 
 angle $\theta$ with the normal to the planes:
\begin{equation}
\cos\theta\ {dI_{\nu}\over d\tau_{\nu}}\  =\ I_{\nu}\ -\ S_{\nu},
\label{eq-transferbis}
\end{equation}
	
The formal solution of this equation is: 
\begin{equation}
\left[I_{\nu}\ \exp \left(-{\tau_\nu\over \mu}\right)\right]^
{\tau_{\nu 2}}_{\tau_{\nu 1}}
\ =\ -\int^{\tau_{\nu 2}}_{\tau_{\nu 1}}S_{\nu}\ \exp 
\left(-{\tau_\nu\over \mu}\right){d\tau_\nu\over \mu}
\end{equation}
where $\mu = \cos\theta$.

	We can apply this solution to different cases, according to the 
boundary conditions.

\subsubsection{for a stellar atmosphere}

In this case the boundary conditions are:

\noindent$\bullet$\quad  no radiation incident on the surface, or 
$I_\nu(\mu<0, \ \tau_{\nu}=0)=0$, i.e.:
\begin{equation}
I_\nu (\mu <0,\tau_{\nu}) = \int_0^{\tau_{\nu}} {S_{\nu}\over |\mu|} 
\exp \left(-{(t-\tau_\nu\over \mu}\right) \ dt,
\end{equation}

\noindent$\bullet$\quad the radiation at $\tau_\nu = \infty$ remains 
finite, i.e.:
\begin{equation}
I_\nu (\mu >0,\tau_{\nu}) = \int^0_{\tau_{\nu}} {S_{\nu}\over |\mu|} 
\exp \left(-{(t-\tau_\nu\over \mu}\right) \ dt \ ,
\end{equation}
and the intensity at the surface is: \index{intensity}						
\begin{equation}
I_\nu (\tau_\nu=0) = \int_0^\infty S_{\nu}
\exp \left(-{\tau_\nu\over \mu}\right) \ {d\tau_\nu\over \mu}.
\label{int_sortant}
\end{equation}
One gets also the flux emerging from the surface: \index{flux}
\begin{eqnarray}
F_\nu (\mu >0) &=& 2 \pi \int_0^1 \mu \, d \mu \int_0^\infty S_{\nu}
\exp \left(-{\tau_\nu\over \mu}\right) \ {d\tau_\nu\over \mu}
\nonumber \\ 
& = & 2 \pi \int_0^\infty S_{\nu} \, {\cal E}_2(\tau_{\nu}) d\tau_{\nu} \, 
,
\label{flux_sortant}
\end{eqnarray}
where ${\cal E}_n(x)$ is the order $n$ integro-exponential:    
${\cal E}_n(x)  =\int_0^\infty exp(-ux) u^{-n} du$.
	
Although Eqs. \ref{int_sortant} and \ref{flux_sortant} are only formal 
solutions which require to know 
the variation of $S_\nu$ as a function of depth at all frequencies, they 
can be of some help to understand intuitively two observations. 

	First we see that the intensity \index{intensity} is approximately 
equal to the 
source 
function at $\tau_\nu = \mu$, and that the internal layers below do not 
contribute to 
the radiation. It explains the limb darkening effect at the surface of 
the 
Sun. According to the Eddington-Barbier relation, the source 
function  in the photosphere   \index{source function} is proportional to 
$\tau$, so it 
decreases towards the surface
(but not necessarily in the chromosphere, cf. P. Heinzel's lectures). 
Thus 
when we observe the limb ($\mu \rightarrow 0$), we are seeing layers 
close to 
the ``surface" where $S_\nu$ is small, so $I_\nu$ is also small. When we 
observe 
the 
center of the Sun $(\mu= 1)$, we are seeing deeper layers where $S_\nu$ is 
large, 
and $I_\nu$ is large too. So the center of the disk will appear brighter 
than 
the limb.

	Second, in the case of stars, we do not observe the 
intensity, \index{intensity} but 
the flux, \index{flux} which is approximately equal to the source 
function at 
$\tau_\nu=1$. 
Assuming again that $S_\nu$ decreases with height in the 
photosphere, we 
can understand why we see absorption lines \index{absorption line} in the stellar 
spectrum. 
The absorption coefficient $\chi_\nu$ \index{absorption coefficient}  
in a line is larger 
than in 
the surrounding continuum, so we are seeing in a line the layers 
close to 
surface, and in the surrounding continuum the deeper 
layers: the lines are then in absorption. \index{absorption line} The 
effect is inverted if the source function \index{source function} 
increases with the height, as 
it 
is the case for lines formed in the chromosphere, or in extended 
envelopes 
of stars: the lines are then in emission. \index{emission line}

\subsubsection{for a homogeneous slab of finite thickness}

One defines an optical thickness \index{optical thickness} increasing 
towards  the observer, 
$d\tau_\nu = \chi_\nu \ d\ell$.
 Since the medium is assumed homogeneous, the optical thickness of the 
 slab is 
$T_\nu = \int \chi_\nu \ d\ellÊ= \chi_\nu \ H$
in the direction of the normal, where $H$ is the geometrical thickness.

	The boundary conditions are now:
\begin{itemize}
\item incident intensity \index{intensity} $I_\nu (\tau_\nu = 0) = I_{\nu 
0}$,
  
\item no incident intensity at $\tau_\nu = T_\nu$. \index{intensity}
\end{itemize}

	The solution of the transfer equation  \index{transfer equation}
is thus (for $S_\nu =$ const.):
\begin{equation}
I_\nu(T_\nu, \mu) 
= I_{\nu 0}\ \exp \left(-{T_\nu\over \mu}\right) +
 S_{\nu} \left[1 - \exp \left(-{T_\nu\over \mu}\right) \right] \ .
\end{equation}
If $T_\nu < 1$, the slab is called ``optically thin"; 
if $T_\nu > 1$ it is called ``optically thick".

	There are three interesting cases:
\begin{itemize}

\item 1.  a non emissive layer (a cold cloud in front of 
an intense source):
\begin{equation}
I_\nu (T_\nu, \mu) = I_{\nu 0} \ \exp  \left(-{T_\nu\over \mu}\right)
\end{equation}

\item 2. an optically thin layer with $T_\nu \ll 1$ 
(assuming that $\mu$ is not $\ll 1$): 
\begin{eqnarray}
I_\nu(T_\nu, \mu) & = & I_{\nu 0} \ \exp  \left(-{T_\nu\over \mu}\right)
+ {S_\nu\ T_\nu\over \mu} \nonumber \\
& = & I_{\nu 0} \ \exp  \left(-{T_\nu\over \mu}\right) + {H \eta_\nu \over \mu}
\label{eq-I-opt-thin}
\end{eqnarray}

\item 3. an optically thick layer with $T_\nu  \gg 1 $:
\begin{equation}
I_\nu(T_\nu, \mu) \sim   S_\nu .
\end{equation}	
\end{itemize}

	For a finite slab with no incident radiation one can show that 
$F_\nu=2\pi S_\nu[0.5-{\cal E}_3(T_{\nu})]$, and deduce that: 
\begin{itemize}
\item 1. in the optically thin case: $F_\nu=2\pi S_\nu T_{\nu}= 2\pi\eta_\nu H$,
 or: $L_\nu = 4\pi\eta_\nu\times$ Volume (where $L_{\nu}$ is the total 
power
 emitted by the slab), 
\item 2.  in the optically thick case: $F_\nu=\pi S_\nu$, or $L_\nu=\pi 
S_\nu\times$ Surface.
\end{itemize}			       
	To summarize, in the optically thin case, one ``sees" the 
emissivity, \index{emissivity} and 
the power is proportional to the volume. In the optically thick case, 
one 
``sees" the source function, \index{source function} and the power is proportional to the 
surface.

Finally note that in a purely diffusing medium, the same equations hold, 
replacing $S_\nu$ by $J_\nu$, and $\tau_\nu$ by the diffusion coefficient.
 \index{diffusion coefficient}
So one gets in particular for an optically thin medium (i.e. 
$T_\nu^{dif}\ll 1$) with no incident radiation on the line of sight:
\begin{equation}
I_{\nu} = {J_\nu T_\nu^{dif}\over \mu}\ .
\label{eq-transfer-diffter}
\end{equation}
This equation applies to the continuum of the solar corona in the visible 
range (the K corona). 
For a medium which is absorbing and diffusing, and optically thin both for 
diffusion and for absorption, but where the diffusion coefficient 
 \index{diffusion coefficient} is 
negligible compared to the absorption coefficient, \index{absorption coefficient} 
the solution of the 
transfer equation becomes:  \index{transfer equation}
\begin{equation}
I_{\nu} = {J_\nu T_\nu^{dif}\ + H \eta_\nu\over \mu}\ .
\label{eq-transfer-diff4}
\end{equation}
This equation applies to the visible lines emitted by the solar corona 
(cf. later)

\section{Local Thermodynamical Equilibrium (LTE)}  \index{LTE (local thermodynamical equilibrium)} \index{LTE (local thermodynamical equilibrium)}

\subsection{RECALLING THE Laws of Thermodynamical Equilibrium (TE)} 
\index{thermodynamical equilibrium}

The {\it thermodynamical equilibrium} is the stationary state of an 
ensemble of interacting
particles and photons which should be achieved in an infinitely thick 
medium (called 
a 
``Black Body") after an infinite time. \index{black body} Photons and 
particles  have then the 
most 
probable energy distribution, which corresponds to 
microreversibility 
of all processes. \index{microreversibility} For instance, there are as 
many radiative 
(resp. collisional) 
excitations  \index{excitation} from the level A to the level B of an atom, as radiative 
(resp. collisional) deexcitations   \index{deexcitation} from the level B to the level A per 
unit time.

\subsubsection{Energy distribution of photons: the Planck law} 
\index{Planck}
It writes:	
\begin{equation}
I_\nu \equiv B_\nu = {2 h \nu^3 \over c^2}
\left[\exp \left({h \nu \over kT} \right) - 1 \right]^{-1},
\end{equation}
\begin{itemize}	
\item $h$ :  Planck  constant = 6.6262 10$^{-27}$  erg s, 
\item $k$ : Boltzmann constant = 1.3806 10$^{-16}$ erg K$^{-1}$
\item $T$ : temperature.
\end{itemize}
Caution: it can also be expressed in units of wavelength, and it writes 
then:
\begin{equation}
B_\lambda = {2 h c^2 \over \lambda^5}
\left[\exp \left({h c \over \lambda kT} \right) - 1 \right]^{-1},
\end{equation}
which has different shape and position of the maximum.

	The integration over $\nu$ or $\lambda$ gives:
\begin{equation}
B = \int B_\nu \ d \nu = {\sigma T^4 \over \pi} ,
\end{equation}
where $\sigma$ is the Stefan  
constant = 5.6696 10$^{-5}$ erg cm$^{-2}$ s$^{-1}$ K$^{-4}$.
	
	There are two limiting cases :
\begin{itemize}	
\item for $h \nu \ll kT$, the Rayleigh-Jeans law (used in radio-astronomy):
\begin{equation}
B_\nu = {2 h \nu^3 \over c^2} \ {kT \over h \nu} = {2 kT \over \lambda ^2} 
;
\end{equation}

\item for $h\nu \gg kT$, the Wien law (used in the X-ray range):
\begin{equation}
B_\nu = {2 h \nu^3 \over c^2}
\exp \left( - {h \nu \over kT} \right) .
\end{equation}
\end{itemize}

\subsubsection{Energy distribution of particles in non quantified 
levels: \\ the 
Maxwell law} \index{Maxwell law}
It gives the number of particles per unit volume whose velocity 
projected 
on an axis $z$ is between $v_z$ and $v_z + dv_z$:
\begin{equation}
dN_z  = N \left({M \over 2\pi kT} \right)^{1/2}  
\exp \left(-{ M v_z^2 \over 2kT} \right)  dv_z 	\ ,
\label{eq-maxw1}	
\end{equation}
where $N$ is the number of particles per unit volume, and $M$ is the 
mass of the particles.
	This expression is used e.g. to compute the profile of spectral lines 
broadened by thermal Doppler effect (cf. later). \index{thermal broadening}

	Since at TE  \index{thermodynamical equilibrium} 
the velocity distribution is isotropic, the number of 
particles per unit volume whose absolute velocity is between 
$v= (v_x^2 + v_y^2 + v_z^2)^{1/2}$  and $v + dv$  is equal to:
\begin{eqnarray}
dN & = & N \left({M \over 2\pi kT} \right)^{3/2}
\exp \left(- {M (v_x^2 + v_y^2 + v_z^2) \over 2kT} \right) \  dv_x \ 
dv_y \ dv_z \nonumber \\	
& = &  N  \left({M \over 2\pi kT} \right)^{3/2}  
\exp \left(-{ Mv^2 \over 2kT} \right) \ 4\pi v^2 \ dv \ ,
\end{eqnarray}		
and the number of particles per unit volume with energy between 
$E = M v^2/2$ and $E+dE$  is:
\begin{equation}
dN  = N \left({1 \over \pi kT} \right)^{3/2}  
\exp \left(-{E\over kT} \right) 2 \pi \ E^{1/2} \ dE	\ .
\label{eq-maxwell-energy}	
\end{equation}
	This expression is used for instance to compute the rate of 
collisional excitations (cf. later).  \index{excitation} 

	Note that the Maxwellian distribution 
\index{Maxwell law} is normalized 
($ \int dN = N$).

\subsubsection{Energy distribution of particles in quantified levels: \\ 
the 
Boltzmann law} \index{Boltzmann law}
Consider two levels of energy $E_m$ and $E_n$ ($E_m > E_n$), with 
$E_m - E_nÊ=ÊE_{nm}=h\nu_{nm}$, where $\nu_{nm}$ is the frequency of the 
corresponding atomic
transition. The 
number of particles per unit volume in each level (called the 
``level population") is given by:
\begin{equation}
{N_m \over N_n} = {g_m \over g_n} \ \exp \left(-{E_{nm} \over kT}\right)
\end{equation}
where  $g_m$ and $g_n$ are the ``statistical weights" of the levels $m$ and 
$n$ (they 
are given by spectroscopic tables, as well as the energies of the levels).

\subsubsection{Distribution of particles in different ionization 
states: \\ the 
Saha law} \index{Saha law}
It gives the distribution of particles of a given species in a given 
ionization state:
\begin{equation}
{N_{i+1}^1 N_e \over N_i^1} = {2 g_{i+1}^1 \over g_i^1} 
\left({2 \pi m kT \over h^2}\right)^{3/2} \exp \left(-{\chi_i \over 
kT}\right) \ ;
\end{equation}
\begin{itemize}

\item $N_{i+1}^1$ and $N_i^1$ are the numbers of ions on the fundamental 
level of 
states ionized $i+1$ and $i$ times, per unit volume,

\item $N_e$ is number of electrons, per unit volume,

\item $g_{i+1}^1$ and $g_i^1$ are the statistical weights of the 
fundamental level 
of states $i+1$ and $i$, \index{statistical weight}

\item $\chi_i$ is the ionization potential of state $i$ (i.e. the energy 
needed to 
extract an electron from the fundamental level), \index{ionization potential}

\item $m$ is the electron mass (9.1096 10$^{-28}$ g).

\end{itemize}

	Using Boltzmann law, one gets:  \index{Boltzmann law}

\begin{equation}
{N_{i+1} N_e \over N_i} = {2 U_{i+1} \over U_i} 
\left({2 \pi m kT \over h^2}\right)^{3/2} \exp \left(-{\chi_i \over 
kT}\right) \ ;
\end{equation}
\begin{itemize}	

\item $N_i$ and $N_{i+1}$ are the total numbers of ions in states ionized 
$i$ and $i+1$ times, per unit volume,

\item $U_i$ and $U_{i+1}$ are called the ``partition functions": 
\index{partition function}
\begin{equation}
U_i(T) = \sum g_m\ \exp \left(-{E_{1m} \over 
kT}\right) \ .
\end{equation}
				
\end{itemize}

	For small $T$ ($E_{1m}/kT >> 1$),  $U_i(T)=g_1$. For large $T$, the 
summation must be 
made on a large number of levels (depending on the density). 

	One can show (cf. books on statistical mechanics) that 
{\it microreversibility 
is equivalent to Maxwell + Boltzmann + Saha equations}. \index{microreversibility} 

	At TE  \index{thermodynamical equilibrium} 
the optical thickness  \index{optical thickness}  
is infinite and the intensity 
\index{intensity} is isotropic,
 so: 
\begin{equation}
B_{\nu}(T)=I_{\nu}(\tau_{\nu}\rightarrow 
\infty)= S_{\nu}
\end{equation}
which writes: 
\begin{equation}
\eta_{\nu}\ =\ \chi_{\nu}\ B_{\nu}(T).
\end{equation}
	It is the well known {\it  Kirchhoff law}, \index{Kirchhoff law} which 
 is used to 
compute 
the emissivity \index{emissivity} when the absorption coefficient is known (or 
inversely). \index{absorption coefficient} 
		          	       
\subsection{From TE to LTE}  \index{LTE (local thermodynamical equilibrium)}

At TE,  $I_{\nu}$ is equal to $ 
B_{\nu}(T)$ and since $T$ is constant, $ dI_{\nu}/d\tau_{\nu}=0$
and there is no transfer of radiation.  

\begin{figure}
\centerline{\psfig{file=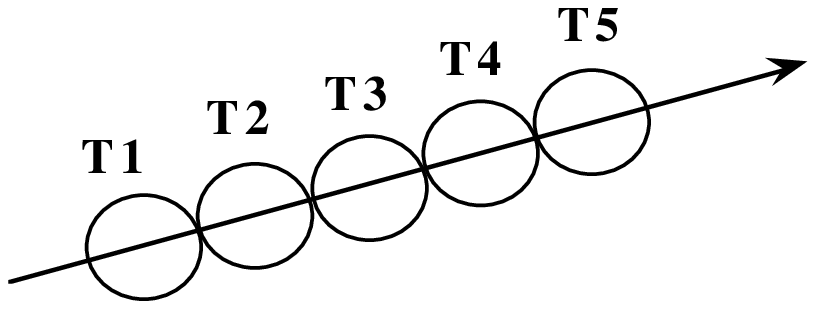,width=0.5\textwidth,angle=0,clip=}}
\caption{Local Thermodynamical Equilibrium.  \index{LTE (local thermodynamical equilibrium)}} 
\label{fig-LTE}
\end{figure}

	Let us now consider a state where {\it $T$ varies in the medium, but 
Maxwell, 
Boltzmann and Saha equations are satisfied locally} (cf. Fig. 
\ref{fig-LTE}).
	In such a state {\it microreversibility is achieved}. \index{microreversibility} 
 It is called {\it 
Local 
Thermodynamical Equilibrium (LTE)}  \index{LTE (local thermodynamical equilibrium)}. We will show later that this 
state is 
reached if the density and/or the optical thickness  \index{optical thickness} are 
large. The source function \index{source function} is then given by the Planck law (but 
not $I_{\nu}$, \index{Planck}
which varies in the medium and is the solution of the transfer 
equation).    

\section{Source function of a spectral line in non-LTE (NLTE)} 
\index{source function} 
\index{NLTE (non-local thermodynamical equilibrium)}

	A spectral line corresponds to the radiative transition 
between two bound 
levels of a given atom. If LTE  \index{LTE (local thermodynamical equilibrium)} is not reached, there is no 
microreversibility  
\index{microreversibility}
between 
the transitions but simply a stationary equilibrium in which the number 
of ALL 
processes populating a level is equal to the number of ALL 
processes 
depopulating this level. The main direct and inverse atomic 
processes 
populating and depopulating a bound level are:

\begin{itemize}

\item collisional excitations and deexcitations  \index{excitation} 
  \index{deexcitation} 

\item radiative excitations and deexcitations

\item collisional ionizations and recombinations \index{recombination}
\index{collisional ionization}

\item radiative ionizations and recombinations.

\end{itemize}

 Several other processes can 
be 
important, such as dielectronic recombinations and auto ionizations, 
Auger 
processes, charge exchanges, but we will not 
consider them here.

\subsection{RADIATIVE EXCITATIONS AND DEEXCITATIONS}  
  \index{deexcitation}  \index{excitation} 

Consider again the two bound levels $m$ and $n$ with populations $N_m$ and 
$N_n$, energies $E_mÊ>ÊE_n$, $E_{nm}
= E_m-E_n$. 

\subsubsection{Einstein probabilities} \index{Einstein probabilities}

	The Einstein probabilities $A_{mn}$, $B_{mn}$ and $B_{nm}$
are defined as follows:

\begin{itemize}

\item the number of radiative spontaneous transitions from level $m$ to 
level $n$ 
per unit volume per second is $N_mÊA_{mn}$. Each corresponds to the 
{\it emission} of a 
photon $h{\nu} = E_{nm}$.

\item the number of radiative transitions from level $n$ to level $m$
  per unit volume per second,
corresponding to the {\it absorption} of a 
photon $h{\nu} = E_{nm}$, is $N_nÊJ_{\nu}B_{nm}$.

\item the number of radiative transitions from level $m$ to level $n$, 
induced by a 
photon $h{\nu} = E_{nm}$, per unit volume per second and leading to 
the emission of a second photon $h{\nu} = E_{nm}$, is $N_mÊJ_{\nu}B_{mn}$.
 It is called {\it ``induced 
emission"}
(discovered by Einstein). Note that $J_{\nu}$ is the mean intensity 
\index{intensity} at  
frequency 
$\nu = E_{nm} /h$. 

\end{itemize}
	
	One can determine the relations between these coefficients in 
TE.  \index{LTE (local thermodynamical equilibrium)} Microreversibility  \index{microreversibility} 
between radiative excitation and deexcitation   \index{excitation} 
  \index{deexcitation} 
gives:
\begin{equation}
N_m(A_{mn}+B_{mn}J_{\nu})\ =\ N_nB_{nm}J_{\nu}
\label{eq-Einstein1}
\end{equation}
Boltzmann equation  \index{Boltzmann law} between levels $m$ and $n$, and the Planck law 
\index{Planck} $J_\nu 
\equiv B_\nu$ lead to:
\begin{equation}
J_{\nu}=\left({A_{mn}\over B_{mn}}\right)
{g_mB_{mn}\over g_nB_{nm}}\left[\exp\left({E_{nm}\over kT}
\right)-1\right]^{-1}
\label{eq-Einstein2}
\end{equation}

	Identifying the terms depending and not 
depending 
on $T$ (as these relations should be verified for any $T$), one gets:
\begin{equation}
g_nB_{nm}=g_mB_{mn},\ \ {\rm and} \ \ {A_{mn}\over B_{mn}}={2h\nu^3\over 
c^2}.
\label{eq-Einstein3}
\end{equation}

	{\it These relations are equally valid in non-LTE}, 
\index{NLTE (non-local thermodynamical equilibrium)}
since $A_{mn}$, $B_{mn}$ and 
$B_{nm}$ 
are atomic 
parameters and do not depend on the physical conditions of the medium.
The $A_{mn}$ are given in spectroscopic tables (or equivalently the 
oscillator strength $f_{nm}$, 
cf. later).
Note that a different definition of the Einstein probabilities is 
 \index{Einstein probabilities}
sometimes used, with the density of radiation instead of the intensity 
in Eq. \ref{eq-Einstein1}.

\subsubsection{Relations of the Einstein probabilities with the absorption 
coefficient, the emissivity, \index{emissivity} and 
the 
source function} \index{source function}  \index{Einstein probabilities}

Spectral lines are broadened by several processes, corresponding to 
an 
absorption profile $\Phi(\nu)$ \index{absorption profile} and an emission 
profile $\Psi(\nu)$, normalized so 
that $\int \Phi(\nu) d\nu=1$ and $\int \Psi(\nu) d\nu=1$.
The emissivity \index{emissivity} then writes:
\begin{equation}
\eta_\nu = {h \nu \over 4 \pi} \ N_m \, A_{mn} \Psi(\nu),
\label{eq-emis-profile}
\end{equation}
and the absorption coefficient: \index{absorption coefficient} 
\begin{equation}
\chi_\nu = {h \nu \over 4 \pi} \ N_n \, B_{nm} \Phi(\nu)
\left(1 - {N_m \ B_{mn} \over N_n \ B_{nm}}\right) \ .
\end{equation}
	It contains a negative term 
of induced 
emission, $(N_mB_{mn})/(N_nB_{nm})=(N_mg_n)/(N_ng_m)$, which can be 
larger than unity 
in the radio range (because the exponential term in the Boltzmann equation 
is close to unity).  \index{Boltzmann law} In this case the medium becomes an amplifier 
instead 
of an absorbant (MASER effect).

	If $\Psi(\nu) = \Phi(\nu)$, one gets: 
\begin{equation}
S={\eta_{\nu}\over \chi_{\nu}}=\left({N_mA_{mn}\over 
N_nB_{nm}}\right)
\left(1-{N_mB_{mn}\over N_nB_{nm}}\right)^{-1}
\label{eq-sourcefunction}
\end{equation}
and the source function \index{source function}  is constant along the line profile.
 
	The absorption and emission profiles are generally the convolution of 
a Gaussian 
and a Lorentzian function, the first being due to thermal and turbulent 
broadening, the latter due to radiative and 
collisional damping. Such a convolution is called ``Voigt profile". The 
emission  and absorption profiles 
 \index{absorption profile} are 
identical in the case of 
coherent scattering, or if on the contrary there 
is complete redistribution of frequencies in the absorption-reemission 
process. It is the case for a majority of lines, in particular for the 
coronal 
lines considered 
later 
in this chapter. Since they correspond to forbidden transitions, 
\index{forbidden line}
radiative damping is 
negligible, and since they are formed in a dilute medium, collisional 
damping is also negligible. So the emission/absorption 
profile reduces to a gaussian 
function. Let us rapidly recall the theory.

If an atom with a transition $ h\nu_0 = E_{nm}$ has 
a velocity projected on the line of sight equal to $v_z$, it
 absorbs or emits photons propagating towards the observer,
shifted by Doppler effect to $\nu=\nu_0(1+v_z/c)$. For an 
assembly of atoms with a Maxwellian distribution of velocities, 
\index{Maxwell law} the 
number of atoms able to absorb or 
emit 
at a frequency $\nu$, in a unit frequency interval, is therefore, 
according to Eq. 
\ref{eq-maxw1}:
\begin{equation}
{dN_z\over dv_z} \ {dv_z\over d\nu} = {c\over \nu_0} N_n \left({M \over 
2\pi kT} 
\right)^{1/2}  
\exp \left(-{ M v_z^2 \over 2kT} \right).
\end{equation}
and one deduces for the absorption/emission profile:
\begin{equation}
\Phi(\nu) = {1 \over \sqrt \pi \Delta \nu_D} \ 
\exp\left[-\left({\nu - \nu_0 \over \Delta \nu_D} \right)^2\right] \ ,
\label{eq-phinu}
\end{equation}
where 
\begin{equation}
\Delta \nu_D = {\nu_0 \over c} \ \left({2 kT \over M}\right)^{1/2}
\label{eq-dopplerwidth}
\end{equation}
is called the ``Doppler width" of the line. \index{Doppler width}
If the gas has also a micro-turbulent velocity $V_{\rm 
turb}$, it should be 
taken into account in the Doppler width: 
\begin{equation}
\Delta \nu_D = {\nu_0 \over c} \ \left({2 kT\over M} + 
V_{\rm turb}^2\right)^{1/2} \ 
\end{equation}
as turbulence corresponds also to a Gaussian distribution of velocities.

The absorption coefficient  \index{absorption coefficient}  
at a frequency $\nu$ thus writes:
\begin{equation}
\chi_\nu ={h \nu_0 \over 4 \pi} \ N_n B_{nm}\, 
{1 \over \sqrt \pi \Delta \nu_D} \ \exp\left[-\left({\nu - \nu_0 \over 
\Delta 
\nu_D} \right)^2\right] \left(1 - {N_m \ B_{mn} \over N_n \ B_{nm}}\right) 
\ 
\end{equation}
and the absorption coefficient at the line center is: \index{absorption coefficient} 
\begin{equation}
\chi_{\nu_0} ={h \nu_0 \over 4 \pi} \ N_nB_{nm} \, 
{1 \over \sqrt \pi \Delta \nu_D}
 \left(1 - {N_m \ B_{mn} \over N_n \ B_{nm}}\right)\ .
\end{equation} 

	Instead of the Einstein probability,  \index{Einstein probabilities}
one often uses {\it the oscillator 
strength} $f_{nm}$ \index{oscillator strength} which is related to 
$B_{nm}$  by:
\begin{equation}
B_{nm}={\pi e^2\over mc} f_{nm} {4\pi\over h\nu}
\label{eq-fnm}
\end{equation}
leading to:
\begin{equation}
\chi_{\nu_0}=N_n{\pi  e^2\over mc} f_{nm} {1 \over \sqrt \pi \Delta \nu_D}
 \left(1 - {N_m \ B_{mn} \over N_n \ B_{nm}}\right)\ .
\label{eq-chi-linecenter}
\end{equation}

	Numerically, $\pi e^2/ mc=0.027$ in CGS,
 $g_nf_{nm} = 1.5\times 10^{-8}g_m A_{mn} \lambda^2$  ($\lambda$
 in micron, $A_{mn}$ in s$^{-1}$), so  $f$ is 
of the order of unity for permitted lines, and $\ll$ 1 for forbidden 
lines.  \index{forbidden line} As an example, for hydrogen:
\begin{equation}
f_{nm}={2^6\over 3\sqrt{3}\pi}{1\over g_n}\left({1\over n^2}-{1\over 
m^2}\right)^{-3}{1\over m^3}{1\over n^3}\ g
\end{equation}
where $g$ is the so-called ``Gaunt factor", of the order of unity.

\subsection{COLLISIONAL EXCITATIONS AND DEEXCITATIONS} 
  \index{deexcitation}   \index{excitation} 

	In passing close to an atom, a perturber P, having a kinetic energy
$E_{\rm kin}$, can loose kinetic energy in exciting
an atom from the bound level $n$ to the bound level $m$. The kinetic energy
$E'_{\rm kin}$ of the perturber after the collision is equal to $E_{\rm 
kin}
  - E_{nm}$. In the inverse process, the 
perturber gains the energy $E_{nm}$ in deexciting the atom from the level 
$m$ to the level $n$.

	One defines the ``collisional excitation rate" \index{excitation}
 $C_{nm}$ and the
 ``collisional 
deexcitation rate" $C_{mn}$ as follows:   \index{deexcitation} 
\begin{itemize} 
 \item the number of collisional excitations per second per unit volume is 
$N_P \ N_n \ C_{nm}$
 \item the number of collisional deexcitations per second per unit volume 
is  $N_P \ N_m \ C_{mn}$, \index{deexcitation}
\end{itemize}
where $N_P$ is the number density of the perturbers.

	The perturbers are most often electrons, except in cold 
interstellar 
medium, where they can be hydrogen atoms or molecules. Let us 
call $\sigma_{mn} (v)$ 
(resp.. $\sigma_{nm} (v)$) the cross section for a collisional 
deexcitation  (resp.. a 
collisional excitation), $v$ being the relative velocity of the 
interacting 
particles. Generally the mean kinetic energy of 
different types of particles (in other words their temperature) is 
the 
same. The velocity of the electrons is thus  much larger than the 
velocity 
of the (more massive) ions. If the perturbers are electrons, the 
relative 
velocity $v$ reduces therefore to the velocity of the electrons. 
According 
to the definition of $C_{mn}$  and $C_{nm}$, one gets:
\[
C_{mn} \ {\rm (respt.} C_{nm})  = <\sigma_{mn} v >  \ {\rm (respt.}
 <\sigma_{nm} v >) \ {\rm s}^{-1} \, {\rm cm}^3\, ,
\]	
where the average is taken over the velocity distribution of the 
perturbers, i.e. the 
Maxwellian distribution 
of the electrons: \index{Maxwell law}
\begin{equation}
C_{mn}(T_e) Ê =  \left({m \over 2\pi kT_e} \right)^{3/2}  \int_0^\infty
\sigma_{mn} (v) \exp \left(-{ mv^2 \over 2kT_e} \right) \ 4\pi v^3 \ dv \ 
, 
\end{equation}
\begin{equation}
C_{nm}(T_e) Ê =  \left({m \over 2\pi kT_e} \right)^{3/2}  
\int_{E_{mn}}^\infty
\sigma_{nm} (v) \exp \left(-{mv^2 \over 2kT_e} \right) \ 4\pi v^3 \ dv \ .
\end{equation}
where $T_e$ is the electron temperature (generally equal to the 
ion temperature). One should 
note 
that for $C_{nm}(T_e)$ the integration is performed from $E_{mn}$ to  
$\infty$, since the perturber
 must have a 
kinetic energy larger than $E_{nm}$ to be able to excite the transition. 
The 
cross sections are determined experimentally and/or theoretically 
as 
functions of $v$, then integrated over the Maxwellian distribution, 
\index{Maxwell law} and 
$C_{mn}(T_e)$ is tabulated.

	As for the radiative rates, one can determine the relations 
between $C_{nm}$ and $C_{mn}$ in TE:   \index{LTE (local thermodynamical equilibrium)}
\begin{itemize} 
 \item 1. microreversibility  \index{microreversibility} between collisional excitation and 
deexcitation   \index{excitation}   \index{deexcitation} 
gives:
\begin{equation}
N_n \ N_e \ C_{nm} = N_m \ N_e \ C_{mn} \ ,		
\end{equation}
 \item
2.  Boltzmann equation between levels $m$ and $n$ gives therefore: 
 \index{Boltzmann law}
\begin{equation}
 C_{nm} = {g_m \over g_n} \ C_{mn}  \exp \left(-{ E_{nm} \over kT_e} 
\right)
\label{eq-excitation}
\end{equation}
\end{itemize}
where $T_e$ is the electron temperature and $N_e$ the electron number 
density.  This relation is also valid in non-LTE. 
\index{NLTE (non-local thermodynamical equilibrium)}

	Numerically one has:
\begin{equation}
 C_{mn} \ = \ {8.63 \,Ê10^{-6} \over T_e^{1/2}} \ {\Omega_{mn}\over g_m}\ 
\ 
 {\rm s}^{-1} \,  {\rm cm}^3
\label{eq-force-coll}		
\end{equation} 
where $\Omega_{mn}$ is a dimensionless ``collision strength", given in 
different tables and articles. \index{collision strength}

\subsection{PHOTOIONIZATIONS AND RADIATIVE RECOMBINATIONS} 
\index{photoionization} \index{recombination}

A photon with an energy $h \nu$ larger than the energy required to extract 
an electron from a bound level $n$, $\chi_n$, can ionize an atom in 
the state 
$n$, ejecting an electron with a kinetic energy $E_{\rm kin}$. This is 
a ``bound-free" 
process, because the energy of the ejected electron is not quantified, 
while the excitation   \index{excitation} between two 
bound 
levels is a ``bound-bound" process. It is called 
{\it photoionization} \index{photoionization} 
and corresponds to the {\it absorption} of the photon $h \nu$. 
Photoionization 
can 
take place either from outer valence shells, or  from inner shells, 
when 
the ionizing photons have X-ray energies. The inverse process, 
corresponding to the {\it emission} of a photon $h \nu$, is the {\it 
radiative 
recombination}.  \index{recombination} 
Finally, as for bound-bound processes, there is also 
stimulated recombination  \index{recombination} induced by an incident photon.

Note:  we do not discuss here {\it dielectronic recombinations} and 
the inverse 
process, 
{\it autoionizations}. Autoionizations are 
generally unimportant.
	
	One defines the rate of photoionization \index{photoionization} 
per unit volume per 
second from level $n$ as $N_n \ P_n$, and the rate of radiative 
recombinations  \index{recombination}
onto level $n$ 
(which can include dielectronic recombinations) as $N_e  N_{i+1} 
\alpha_n$. 
$\alpha_n$ is called the {\it recombination coefficient}. 
\index{recombination coefficient}

	The absorption coefficient \index{absorption coefficient}  
corresponding to a photoionization \index{photoionization} 
from the 
level $n$ can be computed from quantum mechanics, or determined 
experimentally. For hydrogenic ions it writes:
\begin{eqnarray}
\chi_{\nu n} &=& N_n {2^4\, e^2 c^2 R^2 Z^4 g\over 3^{3/2} \, mc} \,  
{1 \over n^5} \, {1 \over \nu^3}\left[1 - \exp \left(-{h \nu \over k 
T_e} \right) \right]  \, \nonumber \\ 
& = & 2.815 \, 10^{29}\, N_n Z^4 \ g{1 \over n^5} \, {1 \over \nu^3}
\left[1 - \exp \left(-{h \nu \over k 
T_e} \right) \right] \, \ {\rm cm}^{-1} \ ,
\end{eqnarray}
where $g$ is a mean ``Gaunt factor", of order unity (for example it 
is equal 0.9 for photoionization \index{photoionization} from level 2). Note that the 
absorption coefficient \index{absorption coefficient} 
is proportional to $\nu^{-3}$. The dependence in 
$\nu$ is generally different for non hydrogenic ions. The 
factor $\left[1 - \exp \left(-{h \nu \over k 
T_e} \right) \right]$ takes into account induced recombinations 
 \index{recombination} (for the 
demonstration, see Mihalas, 1978). 

$\chi_{\nu n}$  is equal to zero for $h \nu < \chi_n$, which is called the 
{\it photoionization edge}.  
One uses currently also $a_{\nu n} = \chi_{\nu n}/N_n$, called the {\it 
photoionization \index{photoionization} 
cross section}. Its value at the edge, $a_{\nu_0\ n}$, writes 
for 
hydrogenic ions (not taking into account induced recombinations):
 \index{recombination}
\begin{equation}
a_{\nu_0\ n} \ = \  7.906 \, 10^{-18} \, {ng \over Z^2} \   {\rm cm}^2 \ 
.		
\end{equation} 
This coefficient is sometimes extended to non hydrogenic ions by defining 
an 
``effective" $Z$,  $Z+s$, $s$ being of the order of unity, but it is 
better 
to use 
results from quantum-mechanical computations now available in several 
databases, such as obtained by the OPACITY project or the CHIANTI database. 

Using the Kirchhoff  \index{Kirchhoff law} 
law $\eta_{\nu n} = \chi_{\nu n} B_\nu(T)$  and the 
laws of 
TE  \index{LTE (local thermodynamical equilibrium)} one can compute the emissivity \index{emissivity} corresponding to the 
recombination 
process.  \index{recombination}  For 
 hydrogen, it gives:
\begin{equation}
\eta_\nu=N_eN_{H^+} {C\over T^{3/2}}{g\over n^3}{h\over 4\pi}\exp 
\left({\chi_n-h \nu \over kT} \right) 		
\end{equation} 
where $ C=3.26\ 10^{-6}$ in CGS.

	We can now write the photoionization \index{photoionization} 
 rate from a level $n$
\begin{equation}
P_n = \int_{\nu_n}^\infty \chi_{\nu n} \left[1 - \exp \left(-{h \nu \over 
k 
T_e} \right) \right] {4 \pi \over h \nu} \, J_\nu \, d\nu  
\end{equation}
which gives, for hydrogen,
\begin{equation}
P_n= 2.815\, 10^{29}\, N_n {4 \pi \over h} {g \over n^5}
\int_{\nu_n}^\infty \left[1 - \exp \left(-{h \nu \over k 
T_e} \right) \right] { J_\nu \over \nu^4}  \, d\nu \ .
{\rm cm}^{-3} \,  {\rm s}^{-1} \ ,
\end{equation}
The recombination rate onto level $n$ is:  \index{recombination}
\begin{equation}
N_e \ N_{i+1} \ \alpha_n \ = {4 \pi \over h} \int_{\nu_n}^\infty 
{\eta_{\nu n} \over 
\nu} \, d\nu \; {\rm cm}^{-3} \,  {\rm s}^{-1} \ ,
\end{equation} 
which gives for hydrogen:
\begin{equation}
\alpha_n \ = \ {C \over T^{3/2}} {g \over n^3} \exp \left({\chi_n \over 
kT} 
\right) \ {\cal E}_1\left({\chi_n \over kT} \right)
\label{eq-recomb}
\end{equation} 		
where ${\cal E}_1$ is the order 1 integro-exponential.  The total 
recombination coefficient \index{recombination coefficient}
is $\alpha_H = \sum \alpha_n$. As shown by 
Eq. \ref{eq-recomb} 
it depends only weakly on 
the temperature. It is roughly given by $\alpha_H = 2 \, 10^{-11} T^{-1/2} 
\,
\Phi\, {\rm cm}^{-3}    {\rm s}^{-1}$,
where $\Phi$ varies from 1 to 3 for $T$ varying from $10^5$ to $10^3$ K.  
These 
expressions are valid for hydrogenic ions with a multiplication by $Z^2$. 
For other ions, published tables give recombination coefficients 
\index{recombination coefficient} as 
functions of the temperature, or absorption coefficients \index{absorption coefficient} 
as functions 
of the photon energy.

\subsection{COLLISIONAL IONIZATIONS AND RECOMBINATIONS}  \index{recombination}
\index{collisional ionization}

An electron  can also be ejected from 
an atom in the energy state $n$, when 
another free electron is passing 
close  to it (or another perturber, but 
electrons are generally more important, as for collisional excitations): 
  \index{excitation} 
this is called a {\it collisional ionization}. \index{collisional ionization}
After the collision the 
first 
electron has lost an amount of kinetic energy equal to the kinetic energy 
of the second electron plus the potential energy $\chi_n$.  The 
conservation of energy 
writes therefore:
\begin{equation}
 E_{\rm kin,1}\ =\ E'_{\rm kin,1} \, + \, E_{\rm kin,2} \, + \, \chi_n ,
\end{equation}
where   $ E_{\rm kin,1}$ and $ E'_{\rm kin,1}$ are the kinetic energies of 
the  perturber before and after the 
collision, and $ E_{\rm kin,2}$ is the kinetic energy of the ejected 
electron.
In the inverse 
process, a free electron induces the recombination  \index{recombination} 
of a second 
electron on the state $n$: this process is called {\it collisional 
recombination}, or {\it 
three body recombination}
(because three particles are interacting).

The rate of collisional ionizations \index{collisional ionization}
from a level $n$, per unit volume and 
time, and  the rate of collisional recombinations  \index{recombination} 
on level $n$, are 
written 
respectively $N_n N_e C_{nc}$  and $N_e^2 N_{i+1} C_{cn}$. Note that with 
this 
definition $C_{nc}$ 
and $C_{cn}$ do not have the same dimension! Like for collisional 
excitations,   \index{excitation} 
the ionization rate is equal to $< \sigma_{nc} v >$, where the average is 
taken 
on the velocity distribution integrated from the ionization edge, and 
$\sigma_{nc}(v)$ is a 
cross section either measured or computed. A general expression valid for 
a 
Maxwellian distribution \index{Maxwell law} and for $kT_e <\chi_i$ is:
\begin{equation}
C_{1n} \approx 10^{-8} \, {\bf o} \, T_e^{1/2} \, \chi_{i(eV)}^{-2} \, 
\exp\left(- {\chi_i \over kT_e} \right)
\end{equation}
where {\bf o} is here the number of optical electrons and $\chi_{i(eV)}$ 
is 
expressed in eV. As usual one finds the 
relation between the direct and inverse coefficients using the TE 
laws:  \index{LTE (local thermodynamical equilibrium)}
\begin{equation}
C_{n1} = C_{1n} \left[ {N_i \over N_e N_{i+1}} \right]_{\rm Saha}.			
\end{equation}

\subsection{STATIONARY EQUILIBRIUM OF AN ATOM (statistical equilibrium 
equations)} \index{statistical equilibrium}

For each level $n$, one can write an equation corresponding to the equality 
of all processes depopulating and populating the level $n$:
\begin{eqnarray}
&\sum_{m\ne n}& { N_m} \left( B_{mn} \int J_\nu \Phi_\nu  d\nu  +  A_{mn} +
N_e  C_{mn}(T_e) \right) \nonumber \\
& & \qquad\qquad\qquad +  N_e N_{i+1}  \alpha_n(T_e) + N_e^2 N_{i+1} 
C_{cn}(T_e)
 \\
=  &\sum_{m\ne n}& {N_n} \left( B_{nm} \int J_\nu \Phi_\nu  d\nu + 
A_{nm} + 
N_e 
C_{nm}(T_e)  + P_n + N_e   C_{nc}(T_e) \right) \nonumber
\end{eqnarray}
where $A_{mn}=0$ for $m<n$. 

These equations can be simplified using the relations between the inverse 
processes shown above. This is generally done replacing also the level 
populations by their ratios to the LTE populations, $b_n$, defined as: 
 \index{LTE (local thermodynamical equilibrium)}
\begin{equation}
N_n=b_n N_{i+1} N_e {g_n \over 2} 
\left({2 \pi m kT \over h^2}\right)^{-3/2} \exp \left({\chi_n \over 
kT}\right) \ ,
\end{equation}
where $\chi_n$ is the ionization potential from level $n$. \index{ionization potential}

This set of equations forms a linear system 
of $n$ equations and $n$ unknown (the level populations), when $T_e$, 
$N_e$, 
$N_{i+1}$ and $ \int J_\nu \Phi_\nu \, d\nu $ are known. If the medium is 
relatively highly 
ionized, and since it is electrically neutral and made mainly of 
hydrogen,  the electrons are provided mainly by hydrogen, and $N_e$ 
is equal roughly to the number of hydrogen nuclei, 
so $N_e \approx 1.2 N_{\rm H^+} \approx 1.2 N_{\rm H}$ (the 
correction 
of  20\% being due to the contribution from helium and heavy 
elements). If 
the medium is weakly ionized, the electrons are provided by 
elements having a low ionization potential, 
 \index{ionization potential}
like C$^+$ and metals, and one cannot 
estimate a priori the value of $N_e$. We will see later how the $N_{i+1}$  
are 
determined. $T_e$ is known through an energy balance equilibrium, and 
$ \int J_\nu \Phi_\nu \,d\nu$ requires the solution of the transfer 
equation.  
So the problem is 
very complex, and involves the sophisticated numerical methods 
described in P. Heinzel's lectures.

	To simplify the discussion we will assume that the atom is 
reduced to two 
bound levels $n$ and $m$. This picture is correct for some forbidden 
lines,   \index{forbidden line} 
such as the H${\rm I}$ 21 cm transition, for the forbidden coronal 
lines 
considered later,   \index{forbidden line} and generally
 when two levels in the ground configuration can be decoupled from the 
resonant 
lines and from the continuum, due to a large difference in potential 
energies. A most frequent case is also that of 3 levels in the ground 
configuration, which are decoupled from the other levels and from the 
continuum. For a two level atom the equilibrium equation reduces 
then to:
\begin{eqnarray}
& N_m& \left( B_{mn} \int J_\nu \Phi_\nu \, d\nu \, + \, A_{mn} +
N_e \, C_{mn}(T_e) \right) = \nonumber \\
 &N_n& \left( B_{nm} \int J_\nu \Phi_\nu \, d\nu \,  + N_e \,
C_{nm}(T_e) \right) \ .
\label{eq-sourcefunctionbis}
\end{eqnarray} 
	Using Eq. (\ref{eq-sourcefunction}) for the source function, one gets: 
\index{source function} 
\begin{equation}
S({\rm line}) = {\int J_\nu \Phi_\nu \, d\nu \, + \epsilon \, B_\nu(T_e) 
\over
1 + \epsilon} \quad {\rm with} \:
\epsilon = {N_e C_{nm} \over A_{mn}}
\left[1- \exp \left(-{E_{nm} \over k T_e}\right)\right] .
\label{eq-sourcefunctionter}
\end{equation} 
$\int J_\nu \Phi_\nu \, d\nu $ is a {\it diffusion term}, as it does not 
correspond 
to any emission or absorption of radiation. From this expression we 
immediately see that {\it if the density is high, $\epsilon$ dominates and 
$S$(line)=$B_\nu(T_e)$. In other 
words, the transition is at LTE}.  \index{LTE (local thermodynamical equilibrium)} We can also note that LTE is reached 
more 
easily when $\int J_\nu \Phi_\nu \, d\nu$ is small (if the medium is 
optically 
thin, for instance). The expression of $S$ is actually more complicated in 
general because
$\Phi(\nu)\ne \Psi(\nu)$.

	In the case of a multi-level atom, it is possible to 
write the 
source function as: \index{source function} 
\begin{equation}
S({\rm line}) = {\int J_\nu \Phi_\nu \, d\nu \, + \epsilon \, 
B_\nu(T_e)\ +\ \epsilon' \over
1 + \epsilon \ +\ \epsilon''},
\end{equation} 
where $ \epsilon'$ and $ \epsilon''$ contain the other transitions, and to 
solve 
the   system by 
iteration. This method was used in the past for a limited number of 
levels, 
but more sophisticated methods converging much more rapidly are used 
presently (cf. P. Heinzel's lectures). 

\subsection{REMARKS CONCERNING THE NOTION OF TEMPERATURES IN A 
NON-LTE MEDIUM}
\index{NLTE (non-local thermodynamical equilibrium)}

In LTE  \index{LTE (local thermodynamical equilibrium)} a unique temperature is sufficient to describe the medium. 
This is not the case in non-LTE, so it is common to use several 
temperatures, according to the 
process 
considered. The most important temperatures which are introduced are:
\begin{itemize}
\item The {\it ``electron temperature"} $T_e$. \index{temperature! thermal} 
It is the temperature 
given by the mean value of the kinetic energy, assuming a Maxwellian 
distribution; \index{Maxwell law} Eq. \ref{eq-maxwell-energy}  gives thus: 
\begin{equation}
T_e={m<v^2>\over 3k} 
\end{equation} 
where $<v^2>$ is the root mean square velocity. This is the temperature 
intervening in the collisional rates $C_{mn}$ and $C_{nm}$, $C_{cn}$ and 
$C_{nc}$.

\item The {\it ``excitation temperature"} $T_{\rm exc}$, defined by: 
\index{temperature! excitation} 
\begin{equation}
{N_m \over N_n} = {g_m \over g_n} \ \exp \left(-{E_{mn} \over kT_{\rm 
exc}}\right),
\end{equation}
which gives:
\begin{equation}
S_{\nu}=\left({N_mA_{mn}\over N_nB_{nm}}\right)
\left(1-{N_mB_{mn}\over N_nB_{nm}}\right)^{-1}=B_\nu(T_{\rm exc}).
\end{equation}
According to this equation, the absorption coefficient becomes:
\index{absorption coefficient} 
\begin{equation}
\chi_\nu = {h \nu \over 4 \pi} \ N_n \, B_{nm} \Phi(\nu)
\left[1 - \exp \left(-{E_{mn} \over kT_{\rm exc}}\right)\right] \ .
\end{equation}
A negative excitation temperature correspond to a negative absorption 
coefficient, i.e. to a maser or a laser effect.
\item The {\it ``radiation temperature"} $T_r$, defined by: 
\index{temperature! radiation} 
\begin{equation}
I_\nu\ =\ B_\nu(T_r)\ .
\end{equation}
In the Rayleigh-Jeans regime, it leads to:
\begin{equation}
T_r={\lambda^2\over 
2k}\ I_\nu
\end{equation}
and the solution of the transfer equation  \index{transfer equation}
for a finite layer becomes: 
 
\begin{equation}
T_r =T_r(0) \exp\left(-{T_\nu \over \mu} \right) \ +\ T_{\rm exc}
\left[1-\exp \left(-{T_\nu \over \mu} \right) \right].
\end{equation}
This equation is commonly used in the radio range.
\item The {\it ``effective temperature"} $T_{\rm eff}$, defined by: 
\index{temperature! effective} 
\begin{equation}
F=\int F_\nu\ d\nu\ = \ \pi B(T_{\rm eff})\ = \ \sigma T_{\rm eff}^4\ .
\end{equation}
\item The {\it ``color temperature"} $T_c$, defined by:
\index{temperature! color} 
\begin{equation}
{dI_\nu\over d\nu}\ = \ {dB_\nu(T_c)\over d\nu} .
\end{equation}
For instance in a nebula ionized by a hot star, $T_c$  is constant in the 
visible and UV range, and
equal to the surface temperature of the star. Also the color temperature 
of 
the solar corona is that of the photosphere in the visible range (owing to 
diffusion).
\end{itemize}
		
\section{Application to the solar corona}

\subsection{Ionization equilibrium in non-LTE}
\index{NLTE (non-local thermodynamical equilibrium)}

To solve the statistical equilibrium equations \index{statistical equilibrium}
 we need to know the 
number 
of ions in a given ionization state. If for each level the rate of 
collisional ionizations \index{collisional ionization}
is equal to the rate of collisional 
recombinations,  \index{recombination}
Saha's law applies. \index{Saha law} One can show that it occurs for 
large 
values of the density ($N_e \ge 10^{16}$ cm$^{-3}$) which are not reached 
in 
the corona 
and in nebular media. These media are therefore in non-LTE 
\index{NLTE (non-local thermodynamical equilibrium)} for 
ionization 
equilibrium, which is given by the stationarity 
equation:

\medskip
\centerline{{\it rate of (photoionizations + collisional ionizations) 
\index{collisional ionization}
 from all levels}} \index{photoionization} 
 
\centerline{{\it =rate of (radiative + collisional recombinations)  
\index{recombination} on 
all 
levels.}}
\medskip
	Actually the problem is simplified by the fact 
that in these media the 
populations of excited levels are very small, so only ionizations 
from the 
ground level should be taken into account. 

	Other simplifications take place in nebular media and in the 
corona.
\medskip

\noindent a. Nebular media:

	HII regions and planetary nebulae are dilute media photoionized by hot 
stars. Their kinetic or electron temperature \index{temperature! thermal} is relatively low 
($10^4$ K). 
One can show that in these conditions radiative ionizations dominate over 
collisional ionizations, \index{collisional ionization}
 and radiative recombinations dominate over 
collisional recombinations  \index{recombination} in
the 
ionization equilibrium, which reduces to:

\centerline{{\it photoionizations from the ground level}}\index{photoionization} 

\centerline{{\it = radiative recombinations 
on all 
levels.}} \index{recombination}

	The photoionization rate \index{photoionization} depends on the distance, on the
temperature, and on the
luminosity of the hot star. For more details on this subject, one can 
refer 
to Osterbrock's book (1989).
\medskip

\noindent b. Solar corona:

	In the corona $T_e$ is of the order of a few 10$^6$ K, while $T_r$ is
 only 5800 K. Therefore 
photoionizations \index{photoionization} are negligible with respect to 
collisional ionizations. \index{collisional ionization}
On the other hand 
the density is relatively low ($N_e \le  10^{10}$ cm$^{-3}$), and 
collisional 
recombinations  \index{recombination} are negligible with respect to radiative 
recombinations.  \index{recombination} The ionization equilibrium thus writes:

\centerline{{\it collisional ionizations from the ground level}}
\index{collisional ionization}

\centerline{{\it  = radiative 
recombinations  \index{recombination}
on all levels}}

or:
\begin{equation}
N_e\  N_i\ C_{ic}(T_e)\ =   N_e N_{i+1}
\  \alpha_i(T)
\end{equation}
where $\alpha_i(T_e)=\sum_{n=1}^{\infty} 
\alpha_{n}(T_e)$, and $C_{ic}(T_e)$ is 
the rate of ionizations from the 
ground level.

	It can be written approximately:
\begin{equation}
{N_{i+1}\over N_i} = 2\ 10^4 {P_i(T_e)\over n} \left({\chi_H\over 
\chi_i}\right)^2\ \exp\left(-{\chi_i\over kT_e}\right) \left({\chi_i\over 
kT_e}\right)^{-1}
\end{equation}
where
$P_i(T_e)$ is a tabulated function, of the order of unity, 
$n$ is the principal quantum number, and
$\chi_H$, $\chi_i$ are the ionization potentials of hydrogen and of the 
ion $i$.  \index{ionization potential}

\begin{figure}
\centerline{\psfig{file=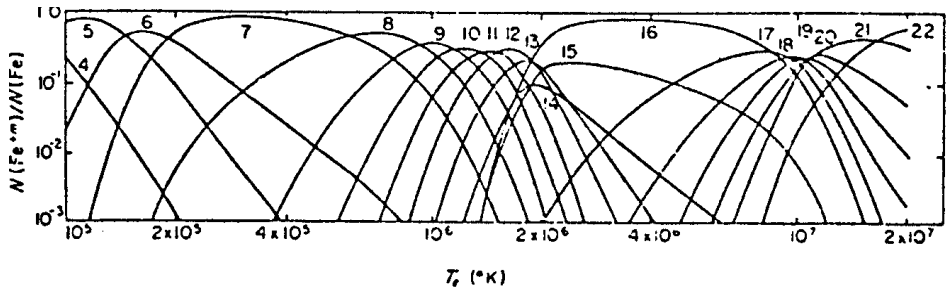,width=1.\textwidth}}
\caption{Coronal equilibrium of iron.}
\label{fig-ion-eq}
\end{figure}

	This equilibrium is called {\it coronal equilibrium}    \index{coronal 
	equilibrium} and holds not 
only in 
the solar corona, but also in the hot interstellar medium, in the hot gas 
 of galaxy 
clusters, in supernovae remnants, etc. Its particularity is that {\it $ 
N_i/\sum N_i$ is a function of only $T_e$}. As an example we show
the ionization equilibrium of iron  on Fig. 
\ref{fig-ion-eq}. We see that for the kinetic temperature 
\index{temperature! thermal} 
of the 
corona, $T_e\sim 10^6$ K, iron should be in the form of highly ionized 
species, Fe$^{+9}$ to 
Fe$^{+13}$ in particular. This figure shows also that the ionic fraction 
is a rapidly varying function of the temperature, implying that if it
can be obtained from the observations, the corresponding
temperature will be accurately determined.

\subsection{THE VISIBLE CORONAL LINES} \index{coronal line}

\begin{figure}
\centerline{\psfig{file=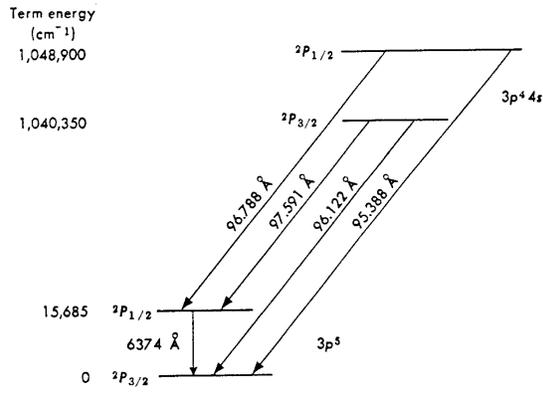,width=0.6\textwidth}}
\caption{Term diagram of FeX showing 4 resonance lines in the EUV range and
the forbidden visible line at 6374 {\AA}.}
\label{fig-niv-Fer}
\end{figure}

	Since the beginning of the century many lines of unknown 
origin have been 
observed in the visible during eclipses. They were attributed to a 
new 
element ``the coronium" (remember the ``nebulium" of planetary 
nebulae!). Grotrian (1939) understood that {\it the intense red line at  
6374{\AA}
  is due to a forbidden transition   \index{forbidden line} in the ground configuration of FeX} 
(cf. Fig. 
\ref{fig-niv-Fer}). 

\begin{figure}
\centerline{\psfig{file=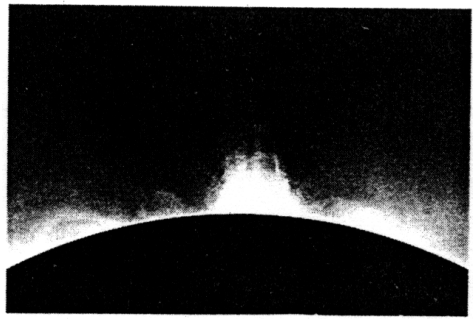,width=0.5\textwidth}}
\caption{An eclipse seen in the forbidden FeXIV 5303{\AA} line}
\label{fig-couronne}
\end{figure}

	Many other lines of highly ionized elements were then 
identified in the visible
coronal spectrum, the most intense being identified  by Edlen in 1942 to 
the forbidden transition in the 3s 
fundamental configuration of Fe$^{+13}$.   \index{forbidden line} 
These observations lead to the discovery of the 
high temperature of the corona. These lines are only visible on the 
limb,  during eclipses, or using a
	coronagraph, as the disk is much too bright in the visible 
	range            (cf. Fig.~\ref{fig-couronne}).

\begin{figure}
\centerline{\psfig{file=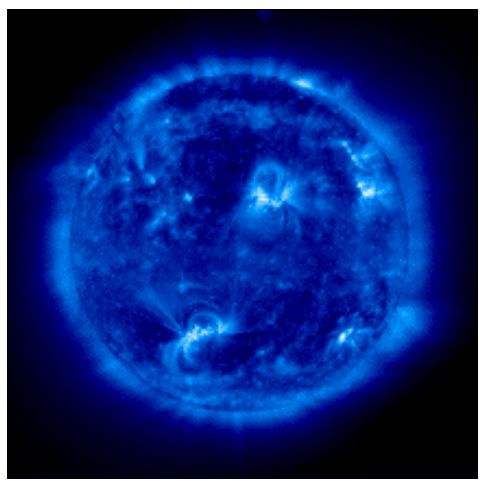,width=0.5\textwidth}}
\caption{An image of the Sun obtained with EIT on board of SOHO in the 
 FeX/XI lines at 171{\AA}.}
\label{fig-SOHO}
\end{figure}

It is not the same for extreme UV and X-ray lines of highly ionized atoms, 
which are emitted only 
by the corona, and can therefore be seen in emission on the disk itself 
because 
there is practically no underlying continuum in these bands. When the extreme 
ultraviolet and the X-ray bands were opened to observation 
with the launch of 
rockets and satellites, 
 a number of permitted 
coronal lines \index{coronal line}  due to resonant transitions were 
discovered. As an 
example Fig. 
\ref{fig-SOHO} displays an image of the Sun obtained with EIT on board of 
SOHO in the FeX/XI lines at 171{\AA}. It shows clearly strong variations of 
the line intensity, due to variations of the temperature and of the 
density in 
the corona (cf. below). 

But here we shall focus on
forbidden visible lines,   \index{forbidden line} 
 because they are more simple to handle than 
resonant lines, and as an example we will determine the intensity 
of the FeXIV 5303{\AA} 
line emitted by the corona. The theory of these lines is established since 
decades,  in particular with the pioneering work of Pottasch (1963). 

\begin{figure}
\centerline{\psfig{file=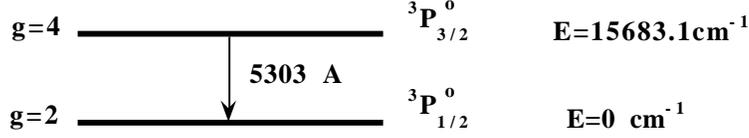,width=0.8\textwidth}}
\caption{The transition FeXIV 5303 {\AA}.}
\label{fig-FeXIV}
\end{figure}

The transition between levels 2 and 1 in the ground configuration of  
Fe$^{+13}$  producing this line is represented on Fig. 
\ref{fig-FeXIV}. One has first to check that it can be
 decoupled from the rest of the atom and from the continuum. This requires 
 to show that 
the rates of radiative and collisional 
 deexcitations   \index{deexcitation} from all upper levels onto level 2, and the rates of 
radiative and collisional 
 recombinations   \index{recombination} 
onto level 2, are negligible compared to  $N_1B_{12} \int 
J_\nu \Phi_\nu 
 d\nu$ and $N_1N_eC_{12}$ (populating level 2), and that the rates 
 of photoionization \index{photoionization} 
 and collisional ionization \index{collisional ionization}
from level 2, and the rates of induced 
radiative and collisional 
 excitations   \index{excitation} to all upper levels from level 2, are negligible compared 
 to $N_2B_{21} \int J_\nu \Phi_\nu 
 d\nu$, $N_2A_{21}$ and $N_2N_eC_{21}$ (depopulating level 2). For 
 instance
 collisional and induced radiative deexcitations   \index{deexcitation} 
from excited levels are 
not 
 important because these levels are not highly populated, owing to their
 large energies, and collisional and 
 radiative excitations   \index{excitation} to
 excited levels are negligible compared to excitations between the 
levels in the ground configuration, owing to the exponential term in the 
excitation rates. 
 Collisional recombinations  \index{recombination} are negligible because of the low density. 
 Actually this is not true for all visible coronal lines, \index{coronal line} and in 
 several cases cascades from highly excited levels, 
 are  important in 
 populating the upper level of the forbidden transition.  \index{forbidden line} 

Let us consider thus a two-level atom in an ionized medium (i.e. 
perturbers are 
electrons), and  
compare the different terms of 
Eq. \ref{eq-sourcefunctionbis}. The atomic data we need are: 
the Einstein coefficient  \index{Einstein probabilities} $A_{21}$ of the transition,
equal to 60 s$^{-1}$, the statistical weights $g_1$ and $g_2$,
 \index{statistical weight}
respectively 2 and 4, and the ``collision strength"  \index{collision strength}
$\Omega_{21}$ (cf. 
Eq. \ref{eq-force-coll} and Fig. \ref{fig-FeXIV}), of the order of unity (it decreases  
 from 3 at $T=10^5$ K to 0.3 at $T=10^7$ K). 

To compute the rate of radiative excitation,   \index{excitation} we need $J_\nu$. 
We will not determine the optical thickness  \index{optical thickness} 
in the continuum at 5303 
{\AA}, as it would require to know the absorption coefficient in the 
continuum at this wavelength, due to free-free process that we have not 
discussed. But we know from observations that the corona is 
optically thin in the optical band, as we see the underlying disk
without any absorption. We infer that the whole corona 
is illuminated by the photospheric radiation field, close to a black 
body of temperature $T_0=5800$K.  Since the line is  narrow 
($\Delta\nu_D\ll \nu_0$), $J_\nu$ is
approximately constant over the line profile. The corona being illuminated 
only 
from one side, one gets: 
\begin{equation}
\int J_\nu \Phi_\nu d\nu = J_\nu =B_\nu(T_r)={ B_\nu(T_0)\over 2} 
\left({R_{\rm sun}\over z+R_{\rm sun}}\right)^2
\end{equation}
where $ R_{\rm sun}$ is the radius of the sun and $z$ the height in the 
corona, and one deduces the 
radiative excitation rate   \index{excitation} (using the relations between the Einstein 
coefficients):  \index{Einstein probabilities}
\begin{equation}
N_1B_{12} J_\nu =N_1\ A_{21}\left[\exp \left({E_{12}\over 
kT_0}\right)-1\right]^{-1}\left({R_{\rm sun}\over z+R_{\rm sun}}\right)^2.
\end{equation}
For the basis of the corona it is equal to $0.9N_1$ s$^{-1}$ cm$^{-3}$, 
and the induced 
deexcitation rate,   \index{deexcitation} 
$N_2{B_{12}\over 2} J_\nu$, is equal to $0.45N_2$ 
s$^{-1}$ cm$^{-3}$, therefore negligible with respect to the 
spontaneous deexcitation rate.  \index{deexcitation} 

 The exponential term in the collisional excitation rate   \index{excitation} $ 
\exp\left(-{E_{12}\over 
kT_e}\right)$ is roughly equal 
 to unity, as $E_{12}\ll kT_e$, so $g_1C_{12}=g_2C_{21}$. 
 According to Eq. \ref{eq-force-coll} one gets thus for the rate of 
collisional
 deexcitation   \index{deexcitation} 
 $N_2N_eC_{21}\sim 2\times 10^{-9}N_2N_e$ s$^{-1}$ cm$^{-3}$ 
 for $T_e=10^6$ K, and the dependence on $T_e$ is small. 
 Comparing this with the radiative deexcitation rate   \index{deexcitation} 
$N_2A_{21}$, we see 
 that above  a critical
density $N_{e}\sim 10^{11}$ cm$^{-3}$, deexcitation   \index{deexcitation} 
would be dominated by collisional processes. It is a very high density, 
and one 
 can admit that {\it deexcitation   \index{deexcitation} is always dominated by radiative 
spontaneous 
 transitions}. For the excitation   \index{excitation} the critical 
 density is much smaller, $\sim  10^{9}$ cm$^{-3}$ at the basis of the 
 corona, and it decreases with 
 height, as the radiative rate is proportional to
 $R_{\rm sun}^2/(z+R_{\rm sun})^{2}$. 
 So  {\it excitation   \index{excitation} is sometimes dominated by radiative
transitions, sometimes by collisional transitions}.

It is important to note here that we have used 3 different temperatures, 
namely the temperature of the underlying blackbody $T_0$, the radiation 
temperature $T_r$, and the electron temperature $T_e$. All are different 
from the excitation temperature $T_{exc}$ (cf. below).
\index{temperature! thermal} \index{temperature! color} 
\index{temperature! radiation} 

 As a rough approximation we can consider
 the corona 
outside the disk as a finite slab of thickness $2R_{\rm 
sun}\sqrt{1-\left(1-{H\over
R_{\rm sun}}\right)^2}\sim 2\sqrt{HR_{\rm sun}}$, 
where $H$ is the scale height of the corona ($< R_{\rm sun}$), with no 
incident intensity on the line 
of sight. The source function \index{source function} in the line is given by Eq. 
\ref{eq-sourcefunctionter}. 
But we do not need this source function \index{source function} 
to determine the line flux, 
\index{flux} 
as we can check that
the corona is optically thin at the center 
of the line (and a fortiori in the wings). Eqs. \ref{eq-dopplerwidth},
 \ref{eq-fnm},
\ref{eq-chi-linecenter}  and the 
relations between the Einstein coefficients, give:  \index{Einstein probabilities}
$\chi_0= 2.3 
10^{-19} N_1$. $N_1$ can be written: ${N_1\over N({\rm Fe}^{+13})}\
{N({\rm Fe}^{+13})\over N({\rm Fe})}\  {N({\rm Fe})\over  N_H}\ {N_H\over 
N_e}\ N_e$, where $N_H$ is the number of hydrogen nuclei per unit volume.  
The iron abundance,   ${N({\rm Fe})\over  N_H}$, is equal to 4 10$^{-5}$, 
and the ratio $ {N_H\over N_e}$ is equal to 0.83 in the corona (owing to 
electrons from helium and heavy elements). Let us make the 
conservative assumption 
that iron is entirely in the 
form of Fe$^{+13}$, and that $N_1$ is much larger than the populations of 
all the other 
levels, i.e. $N_1=N$(Fe$^{+13})$. Both assumptions are reasonable in the 
region emitting the FeXIV line, 
actually. One finds then $\chi_0\sim  10^{-23} N_e$, or $\tau_0\sim 
10^{-23}
 {\cal N}_e$, where  ${\cal N}_e$ is the number of electrons on the 
line of sight per unit surface, in the emission region. Assuming that 
the whole corona is emitting the line (which is clearly an overestimation)
one gets $ {\cal N}_e=<N_e>2\sqrt{HR_{\rm sun}}$, 
where $<N_e>$ is the density averaged on the scale 
height,  $N_e\sim 5\ 10^8$ cm$^{-3}$, and finally one gets for $H\sim 5\ 
10^9$ cm:
$\tau_0\le 
10^{-4} \ll 1$. 

At this stage it is also possible to show that the diffusion coefficient 
 \index{diffusion coefficient}
neglected in all this chapter is indeed negligible compared to the 
absorption coefficient \index{absorption coefficient} in the line. Since diffusion 
is due to Thomson scattering, the diffusion coefficient 
 \index{diffusion coefficient}
is equal to $\sigma_TN_e=0.66\ 10^{-24}N_e$, and we immediately see that 
it is 
smaller than $\chi_0$. However, it is not possible to 
neglect the emissivity \index{emissivity} due to diffusion in the continuum, as it is the 
only emission 
process for the continuum underlying and surrounding the line. According 
to Eq. \ref{eq-transfer-diffter} for an optically thin purely diffusing 
medium, $I_\nu^{dif}=\tau_TJ_\nu \sim\tau_T{ B_\nu(T_0)/2}\sim 0.3\ 
10^{-24} 
{\cal N}_eB_\nu(T_0)\sim 3 10^{-6} B_\nu(T_0)$. This is indeed the intensity 
of the visible continuum observed at the basis of the corona. And according 
to Eq. \ref{eq-transfer-diff4}, the intensity in the line should simply 
be added to the continuum intensity, since the absorption coefficient 
\index{absorption coefficient} 
dominates on the diffusion one. So finally we see that the computation of 
the line intensity
can be completely decoupled from the diffusion process.

The intensity in the line is therefore equal to $\int
 \eta_\nu (z) dz$ 
(as the emissivity \index{emissivity} is strongly dependent on the position, one cannot 
use the homogeneous approximation). It is
 proportional to the emissivity \index{emissivity} at the line center and to 
 the emissivity \index{emissivity} integrated on the line profile (cf. Eqs. 
 \ref{eq-emis-profile} and \ref{eq-phinu}), which writes, using Eq. 
 \ref{eq-sourcefunction}:
\begin{equation}
\eta_{line}={E_{12}\over 4\pi}N_2A_{21}={E_{12}\over 
4\pi}N_1(N_eC_{12}+B_{12}  J_\nu)
\label{eq-emissivity}
\end{equation}
$N_2$ is always smaller than $N_1$, since when excitation   \index{excitation} is dominated by 
collision, ${N_2/N_1}=N_eC_{12}/A_{21}< 1$, and when it is dominated by 
radiation  ${N_2/N_1}=B_{12}J_\nu/A_{21}\sim \exp(-{E_{12}/
kT_0})<1 $. (Note that $T_{exc}$ 
which is defined either by $N_eC_{12}/A_{21}=2\exp(-E_{12}/kT_{exc})$, 
or by $B_{12}J_\nu/A_{21}=2\exp(-E_{12}/kT_{exc})$, is different from 
the previously used temperatures.) 
\index{temperature! excitation} Thus 
 $N_1\sim N({\rm Fe}^{+13})\sim 4\ 10^{-5}N_e{ [N({\rm Fe}^{+13})/ N({\rm 
 Fe})]}$, as assumed before. 

Eq. \ref{eq-emissivity} gives thus, in the two 
extreme cases of $N_e$ much larger and much smaller than the 
critical  density of 10$^9$ cm$^{-3}$: 
\begin{itemize}
\item $N_e >$ 10$^9$ cm$^{-3}$: $\eta_\nu\sim{E_{12}\over 
4\pi}C_{12}\ N_e^2   [N({\rm Fe}^{+13})/ N({\rm Fe})]$
\item $N_e <$ 10$^9$ cm$^{-3}$: $\eta_\nu\sim{E_{12}\over 
4\pi}B_{12} J_\nu\ N_e [N({\rm Fe}^{+13})/ N({\rm Fe})]$.
\end{itemize}
 The first case corresponds to 
dense coronal loops, and the second one is close to the quiet corona. 

Let us assume that the density is larger than the critical one (actually 
it is
 often the case for other forbidden coronal lines).   \index{forbidden line} 
The intensity writes 
 thus:
\begin{equation}
I_{line}\sim 4\ 10^{-5}\ {E_{12}\over 
4\pi} \int {\ N_e^2\  C_{12}(T_e) {N({\rm Fe}^{+13})\over N({\rm Fe})} dz}.
\label{eq-IFeXIV}
\end{equation}

The intensity, and consequently the flux measured at the Earth, is therefore a 
 function of the density and of the temperature in the region where iron 
 is in the form of Fe$^{+13}$. But $N({\rm 
Fe}^{+13})/ N({\rm Fe})$ depends so strongly on the temperature, that one 
considers it  equal to unity in a small range of temperature, and
zero outside. Moreover $ C_{12}(T_e)$ is not strongly dependent on $T_e$. 
One can thus approximate the intensity by
\begin{equation}
I_{line}\sim 4\ 10^{-5}\ {E_{12}\over 4\pi}\ {C_{12}\int E(T_e) dT_e}
\label{eq-IFeXIVbis}
\end{equation}
where $\int E(T_e)dT_e=\int N_e^2dz$ is called the {\it 
emission measure}. \index{emission measure}

 Therefore, 
measuring other visible forbidden coronal lines   \index{forbidden line} 
corresponding to different 
ions should allow to map the emission measure  \index{emission measure}
as a function of the 
temperature. More sophisticated methods involving inversion procedures can 
also be used to avoid the rough approximations made above.

We have focussed here on a forbidden line,   \index{forbidden line} 
but it is clear that the 
same kind 
of computation can be performed for other types of lines, such as resonance 
UV  or X-ray 
lines. The equations giving the population of the upper level of the 
transition would not necessarily have been as simple as in
the particular case considered here, and consequently the dependence of 
the emissivity on the temperature and on the density would have been more 
complicated, but the whole procedure would have been similar.
Then, by comparing  forbidden lines   \index{forbidden line} 
with resonance lines of the 
same 
ion in the extreme UV and in the X-ray, or more generally pairs of lines 
of 
the same ion, one can get powerful diagnostics of 
the temperature and/or of the density in the region where this ion is 
abundant. Besides, the EUV and X-ray lines are observed also on the disk, 
where the forbidden visible lines cannot be observed.   \index{forbidden line} 
For instance SiIX and SiX lines which are density sensitive have been observed 
with the 
coronal diagnostic spectrometer (CDS) on board SOHO and have allowed to 
obtain 
the density and the temperature in active regions on the limb and on the disk
as a function 
of the radius and the position angle. One can guess that many other important results 
will be obtained from the data gathered in this mission.

Finally, one should not forget 
that the simple study of the line profiles using lines of different 
elements allows to determine both the kinetic
temperature and the turbulent velocity, as shown also in this chapter. 

\acknowledgements{
I wish to thank Petr Heinzel for kindly reading the manuscript,
which benefitted from the improvements he suggested.}

{}
\printindex
\end{document}